# Upper critical fields well above 100 T for the superconductor $SmFeAsO_{0.85}F_{0.15}$ with $T_c$ = 46 K

C. Senatore* and R. Flükiger

*Department of Condensed Matter Physics & MaNEP/NCCR, University of Geneva, CH-1211 Geneva 4, Switzerland*

M. Cantoni

*Interdisciplinary Centre for Electron Microscopy (CIME), EPFL, CH-1015 Lausanne, Switzerland*

G. Wu, R. H. Liu and X. H. Chen

*Hefei National Laboratory for Physical Science at Microscale and Department of Physics, University of Science and Technology of China, Hefei, Anhui 230026, China*

**Abstract**

We report specific heat measurements at magnetic fields up to 20 T on the recently discovered superconductor $SmFeAsO_{0.85}F_{0.15}$. The *B*-*T* diagram of a polycrystalline $SmFeAsO_{0.85}F_{0.15}$sample with $T_c$ = 46 K was investigated. The temperature dependence of $B_{c2}$ was extracted from the specific heat curves, the corresponding $B_{c2}(T = 0)$ value derived from the Werthamer-Helfand-Hohenberg formula being 150 T. Based on magnetization measurements up to 9 T, a first estimation of the field dependence of the inductive critical current $J_c$ is given. Evidence for granularity is found. The presence of a peak effect is reported, suggesting a crossover in the vortex dynamics, in analogy to the behaviour observed in high $T_c$ cuprates.

* Corresponding author; Electronic address : carmine.senatore@physics.unige.ch

## 1. Introduction

After the recent discovery of superconductivity in doped quaternary oxypnictides $REFeAsO_{1-x}F_x$, where RE = La, Ce, Pr, Nd, Sm, Gd, this class of compounds has become a subject of considerable experimental and theoretical interest. Soon after the announcement by Kamihara et al. [1] of the synthesis of the novel superconductor $LaFeAsO_{1-x}F_x$ with the transition temperature $T_c$ of 26 K, $T_c$ values up to 55 K were obtained by substituting La with smaller rare earth (RE) elements [2-8]. Ren et al. [9] found that under high pressure, superconductivity in the series $REFeAsO_{1-x}$ can be induced by oxygen vacancies instead of F-doping. The substitution of Oxygen by Fluorine as well as the presence of Oxygen vacancies has the effect of introducing electrons into the REFeAsO system. Kamihara et al. [1] found no trace of superconductivity when substituting $La^{3+}$ with $Ca^{2+}$, i.e. introducing holes into the system. On the other hand, Wen et al. [10] found the evidence of superconductivity through hole doping in the system $La_{1-x}Sr_xFeAsO$ with $T_c$ = 25 K.

Several theoretical works proposed multiband superconductivity arising from Fe-As electronic bands [11-14], suggesting a pairing mechanism of spin fluctuation type. Indeed, the undoped REFeAsO exhibits a spin density wave (SDW) transition at 130-150 K [15-18]. This transition has been detected by various techniques, e.g. electrical resistivity [15], magnetic susceptibility [16], specific heat [17], Hall coefficient [15], and Seebeck coefficient [18]. It was proposed that superconductivity is induced by the suppression of the SDW state (through electron or hole doping), $T_c$ being enhanced after the substitution of La by smaller rare earth atoms, which also leads to a reduced elementary cell [9].

In the present paper we report a study of the magnetic phase diagram *B*-*T* for a polycrystalline $SmFeAsO_{0.85}F_{0.15}$ sample by means of high field specific heat measurements and by magnetization measurements. The sample was prepared in the Hefei National Laboratory [2], and some details of sample preparation are given in Section 2. Specific heat measurements at different magnetic fields up to 20 T allowed the determination of the $T_c$ distribution in the sample by deconvolution (Section 3). These measurements were used for determining the $B_{c2}(T)$ curve up to 20 T. In Section 4 an estimation of the inductive critical current density $J_c$ is given, based on magnetization measurements. The first observation of a peak effect in this family of compounds is reported.

## 2. Sample preparation

The polycrystalline sample with nominal composition $SmFeAsO_{0.85}F_{0.15}$ was synthesized by conventional solid state reaction using SmAs, $SmF_3$, Fe and $Fe_2O_3$ as starting materials [2]. SmAs was obtained by reacting Sm chips and As pieces at 600°C for 3 hours, followed by 5 hours at 900°C. The raw materials were mixed and pressed into pellets, which were wrapped into a Ta foil, sealed in an evacuated quartz tube and annealed at 1160°C for 40 hours. The peaks of the XRD pattern correspond to the tetragonal P4/nmm structure with $a$ = 0.3943 nm and $c$ = 0.8514 nm (Fig. 1a), SmOF being present as impurity phase. As shown by SEM analysis (Fig. 1b), the sample is porous, the conglomerate particle size varying between 5 and 30 μm.

**3. Specific heat measurements**

The specific heat measurements were performed on the polycrystalline sample of $SmFeAsO_{0.85}F_{0.15}$ described above with a home made calorimeter, using the thermal relaxation technique [19]. The sample, with a mass of 44.950 mg, was mounted on a sapphire plate. A Cernox thermometer with four contacts made of phosphor bronze was on the other side of the plate. The whole set-up, including two additional Cernox thermometers and a heater was sealed inside a vacuum chamber. Both thermometers were calibrated in magnetic field. Magnetic fields up to 20 T were generated by a superconducting laboratory magnet with 64 mm bore from Bruker BioSpin. The temperature in the variable temperature insert (VTI) varied between 2 and 50 K.

The variation of $C/T$ versus $T$ in zero magnetic field and at $B$ = 20 T is shown in Fig. 2. The sample exhibits a broad superconducting transition whose midpoint at $B$ = 0 is 45.5 K. The onset of the superconducting transition from AC susceptibility is found around 46 K (inset (a) of Fig. 2).

As shown in the inset (b) of Fig. 2, there is an anomaly in the specific heat curve at temperatures below ~15 K. This behaviour was already observed by Ding et al. [17] and was ascribed to the antiferromagnetic ordering of the $Sm^{3+}$ ions, in analogy to what happens in the electron-doped superconductor $Sm_{2-x}Ce_xCuO_{4-\delta}$ [20].

The measured specific heat $C_{tot}$ above 20 K can be rewritten as

$$C_{tot}(B,T) = C_s(B,T) + C_n(T), \qquad (1)$$

where $C_s(B,T)$ is the superconducting contribution and $C_n(T)$ is the normal background due to phonons and normal electrons. The main difficulty consists in the separation of the singular part $C_s(B,T)$ from the background. For the background we used a quadratic function $C_n(T) = a + bT + cT^2$, and the resulting $C_s(B,T)$ curves are reported in Fig. 3 for $B$ = 0, 6, 10, 14 and 20 T. In the following we characterize the superconducting transition by $T_{mid}$, the midpoint of the superconducting jump, the onset temperature being not well defined. Another point is $T_{max}$ which corresponds to the maximum value of $C/T$ at the superconducting jump (Fig. 3) and is close to the average bulk $T_c$ within the sample. The inhomogeneity of the sample gives rise to a broad specific heat anomaly, which reflects a wide distribution of $T_c$. The calorimetric superconducting transition is broadened when a magnetic field is applied.

The distribution of $T_c$ in the present sample was determined using the previously reported method of analysis based on the deconvolution of the specific heat data [21]. For a distribution of $T_c$ from 0 K to $T_{c,max}$ with a weight function $f(T_c)$, the electronic specific heat $C_e(T)$ and the entropy $S_e(T)$ at a certain temperature $T$ are given by:

$$C_e(T) = \int_T^{T_{c,\max}} f(T_c) C_{es}(T,T_c) dT_c + \int_0^T f(T_c) C_{en}(T,T_c) dT_c\,, \tag{2}$$

$$S_e(T) = \int_T^{T_{c,\max}} f(T_c) S_{es}(T,T_c) dT_c + \int_0^T f(T_c) S_{en}(T,T_c) dT_c\,. \tag{3}$$

The first term of the right side of the equations is the specific heat (eq. 2) and the entropy (eq. 3) of electrons in the superconducting state, while the second part is the specific heat (eq. 2) and the entropy (eq. 3) of electrons in the normal state. Here two assumptions are made:

1) in the sample, there is only one normal-state electronic specific heat coefficient for all the sample parts with different $T_c$:

$$C_{en}(T,T_c) = C_{en}(T) = \gamma T\,; \tag{4}$$

2) the electronic specific heat in the superconducting state can be described using a generalized two-fluid model:

$$C_{es}(T,T_c) = n\gamma T_c (T/T_c)^n\,. \tag{5}$$

We finally obtain

$$\int_0^T f(T_c) dT_c \equiv F(T) = \frac{nS_e(T) - C_e(T)}{(n-1)\gamma T}\,. \tag{6}$$

$F(T)$ is the integral of the distribution between 0 and $T$ and thus represents the fraction of the sample with $T_c \leq T$. The normal-state electronic specific heat coefficient $\gamma$ and the parameter $n$, which in the original Gorter-Casimir model is equal to 3, are chosen in order to obtain

$$F(T_{c,\max}) \equiv \int_0^{T_{c,\max}} f(T_c) dT_c = 1\,.$$

In terms of the notation introduced in eq. (1), eq. (6) can be rewritten as

$$\int_0^T f(T_c) dT_c \equiv F(T) = \frac{nS_e(T) - C_e(T)}{(n-1)\gamma T} = \frac{nS_s(B,T) - C_s(B,T)}{(n-1)\gamma T} + 1\,, \tag{7}$$

where

$$C_s(B,T) = C_e(T) - \gamma T \quad , \quad S_s(B,T) = \int_0^T \frac{C_s(B,T')}{T'} dT'\,. \tag{8}$$

In Fig. 4 we report the $T_c$ distributions measured at $B = 0$ and 10 T. The reduction of the area below the curve at 10 T corresponds to the reduced superconducting volume due to the magnetic flux penetration. Two reasons can be invoked for the increasing width of the $T_c$ distribution with applied magnetic field:

a) the broadening of the upper critical field ($B_{c2}$) line in the $B$-$T$ diagram as an effect of the fluctuations [22-23];
b) the presence of a distribution of $B_{c2}$, due to the intrinsic anisotropy of $B_{c2}$ in the iron based layered superconductors and the random orientation of the grains within the sample.

The temperature dependence of $B_{c2}$ was determined from the points $T_{mid}$ and $T_{max}$ in the measured specific heat curves (Fig. 3), these values being better defined than $T_{onset}$. Compared to other techniques, the calorimetric determination of $B_{c2}$ provides information on the bulk properties of the sample. On the other hand, the results from electrical transport measurements are limited to the percolation paths of the current and the inductive measurements (dc and ac magnetization) are influenced by magnetic shielding effects.

The slope $dB_{c2}/dT|_{Tc}$, extracted from the $B_{c2}(T)$ data for $T_{max}$ is ~5 T/K. The corresponding $B_{c2}(T = 0)$ value derived from the Werthamer-Helfand-Hohenberg (WHH) formula is $0.693T_c \mid dB_{c2}/dT \mid_{Tc} \approx 150$ T, thus exceeding the BCS paramagnetic limit $B_p \cong 1.84T_c \approx 85$ T.

For $LaFeAsO_{1-x}F_x$, Hunte et al. [24] have interpreted the observed upward curvature of $B_{c2}(T)$ as a signature of multiband superconductivity in this family of compounds, which could lead to a significant raise of $B_{c2}(T = 0)$ with respect to the WHH extrapolation. The upper critical field in $SmFeAsO_{0.85}F_{0.15}$ exhibits a very weak temperature dependence close to $T_c$ (Fig. 5). Our data of $B_{c2}(T/T_c)$ and the values measured on $LaFeAsO_{1-x}F_x$ by Hunte et al. [24] and Zhu et al. [25] are reported in Fig. 5.

**4. Inductive measurements**

Magnetization loops $M(B)$ were measured with a vibrating sample magnetometer (VSM). The magnetic field was swept up to 8.8 T and loops were recorded every 2.5 K between 5 K and 55 K.

The experimental $M(B)$ curves can be understood as the superposition of a superconducting contribution and a ferromagnetic background (Fig. 6). The intensity of the signal for the two contributions in the temperature range 5-40 K is comparable.

Measurements performed above the superconducting $T_c$ show a narrow ferromagnetic hysteresis with negligible temperature dependence, probably due to the unreacted Fe and $Fe_2O_3$. The superconducting contribution in the temperature range 5-40 K has been isolated treating the ferromagnetic signal as a temperature independent background. The resulting loops are reported in Fig. 7 for $T$ = 5 K (a) and 25 K (b).

In presence of flux pinning the sign of the superconducting magnetization depends on the direction of the field sweep, thus resulting in a hysteresis loop between the ascending and the descending branches of the magnetization. The critical current density $J_c$ can be obtained from the irreversible magnetization $\Delta M = ( M^+ - M^- ) / 2$ where $M^+$ ($M^-$) is the branch of the magnetization for $dB_a/dt < 0$ ($dB_a/dt > 0$). $B_a$ is the applied field, swept at a constant rate: in our case $| dB_a/dt |$ = 2 T/min. The value $\Delta M$ is proportional to $J_c$ times the length scale of the current flow. The low value of the superconducting magnetic moment for the examined sample suggests the presence of “weak links”, i.e. the current does not circulate through the entire sample. Evidence for electromagnetic granularity was already reported in polycrystalline $LaFeAsO_{1-x}F_x$ [26]. In our case, this was confirmed from the fact that the irreversible magnetization $\Delta M$ does not change reducing the sample size, thus indicating that the current carrying length scale is smaller than the sample dimensions. A further confirmation comes from the double transition present the AC susceptibility curve (Fig 8).

In analogy to high $T_c$ compounds [27], the observed granular behaviour can be ascribed to non-superconducting grain boundaries, to local inhomogeneity or grain misalignment. A possibility is the loss of Fluorine at the grain boundaries as a consequence of the heat treatment performed under vacuum.

In order to calculate the critical current density $J_c(B,T)$ we assumed the current flowing only within the grains. The Bean formula used to calculate $J_c$ from the measured irreversible magnetic moment $\Delta m$ is

$$J_c = 3\frac{\Delta M}{\langle R \rangle} = 3\frac{\Delta m}{V\langle R \rangle}. \qquad (9)$$

No data about the real current length scale $\langle R \rangle$ is known at present, and the determination of $J_c$ is thus uncertain. In the present case, an average conglomerate particle size of $\langle R \rangle \approx 10$ μm can be estimated from SEM measurements, the value of $J_c$ obtained by setting this value in eq. (9) can thus only be considered as a lower limit. Substantially higher $J_c$ values can be obtained when considering decoupled grains inside the conglomerate particles. We have found from FE-SEM analysis that the grain size varies from 50-100 nm to 2-3 μm (Fig. 9). Therefore, an upper limit for the intragrain $J_c$ value can be estimated if we assume a grain size of ~0.1 μm, which is roughly of the same order of magnitude as $Nb_3Sn$, $V_3Ga$, NbTi and $MgB_2$.

Fig. 10 shows the inductive $J_c$ values from 5 to 40 K plotted versus the magnetic field. The $J_c(B)$ values have been represented for both, $\langle R \rangle$ = 0.1 and 10 μm, a more precise answer will be given once the correct length scale of these compounds will be determined.

### *4.1 Peak effect and B-T diagram*

As shown in Fig. 7 (b) and 8, the irreversible magnetization and thus the $J_c(B)$ curves exhibit the peak effect for $T > 10$ K. In many cuprates [28-31] and in some low $T_c$ superconductors such as $Nb_3Sn$ [32], $CeRu_2$ and $NbSe_2$ [33], $V_3Si$ [34] and $MgB_2$ [35-36] the magnetization peak effect has been observed and several models have been proposed to explain the increased pinning as the magnetic field increases to the peak field $B_{peak}$.

In the case of low $T_c$ superconductors, the peak effect appears approaching the $B_{c2}$ line. Close to $B_{c2}$ and thus to the vortex melting, thermal fluctuations are responsible for the softening of the shear modulus in the quasi-ordered Bragg glass state of the vortex matter [37]. This leads to a better accommodation of the vortices on the pinning centres, with the consequent formation of a highly disordered vortex glass corresponding to a sharp increase in the critical current density.

In the highly anisotropic high $T_c$ cuprates, the mixed state can be described in terms of 2D vortices confined in the $CuO_2$ planes and coupled through the separating layers by Josephson and electromagnetic coupling [38-39]. The characteristic length of electromagnetic interaction is the in-plane penetration length $\lambda_{ab}$. In the case of Josephson coupling, the characteristic length is given by $\lambda_J = \Gamma d$ where $\Gamma$ is the anisotropy value and $d$ is the distance between consecutive $CuO_2$ planes. In that respect, the magnetization peak effect appears in the same region of the $B$-$T$ diagram corresponding to the transition from a lattice of 3D flux lines to a lattice of 2D pancake vortices and it results

$$B_{peak} = \frac{\Phi_0}{\ell^2}, \tag{10}$$

$\ell$ being the shorter length between $\lambda_{ab}$ and $\lambda_J$ [30,39-40]. In optimally doped YBCO with $\Gamma < 10$, the peak effect is associated to a crossover from elastic to plastic creep in the vortex motion [31].

In $SmFeAsO_{0.85}F_{0.15}$ the peak in magnetization appears for values of the applied magnetic field much lower than $B_{c2}$. Our scenario is in principle similar to what is proposed for high $T_c$ superconductors. Due to the layered structure of the quaternary oxypnictides, an anisotropic behaviour of the electronic properties is expected. For $LaFeAsO_{1-x}F_x$ Singh et al. [41] predicted a resistivity anisotropy $\rho_c/\rho_{ab} \sim 15$. Therefore the observed magnetization peak effect can be ascribed to a 2D-3D crossover with consequent enhanced vortex pinning. Due to the random orientation of the grains within the sample, this crossover holds only in the grains whose *ab*-plane is not parallel to the applied magnetic field. Recent muon spin rotation experiments yield a penetration length $\lambda_{ab}(T = 0) \approx 200$ nm [42]. Using the expression for the peak field in eq. (10), we find a crossover length spanning between 20 and 50 nm in the temperature range 10–35 K, suggesting that Josephson interactions have to be taken into account.

In Fig. 11 we show the *B*-*T* phase diagram for $SmFeAsO_{0.85}F_{0.15}$. The maximum of the magnetization peak $B_{peak}$ reveals the crossover in the dimensionality of the vortex lattice. The irreversibility field $B_{irr}$ was determined from the extrapolation of the Kramer plot, i.e. $\Delta M^{0.5}B^{0.25}$ versus *B*, at different temperatures. The open squares in Fig. 11 correspond to the separation of the zero field cooled (ZFC) and the field cooled (FC) $M(T)$ curves, measured at $B$ = 1 and 2 T. The irreversibility line follows a $[1-(T/T_c)]^2$ law, as predicted for the melting line in the collective pinning theory [43]. The upper critical field $B_{c2}$ corresponds to the maximum of the superconducting jump in the specific heat measurements.

**5. Conclusions**

The temperature dependence of the upper critical field $B_{c2}$ for the recently discovered $SmFeAsO_{0.85}F_{0.15}$ superconductor was determined from specific heat measurements up to 20 T. The high value of the slope $| dB_{c2}/dT |_{Tc} \approx 5$ T/K suggests an extrapolated $B_{c2}(T = 0)$ around 150 T, using the WHH formula.

We propose a range of values for the critical current density $J_c$, extracted from magnetization measurements, taking into account different length scales, ranging between 10 and 0.1 μm. The granular behaviour of our sample renders the correct estimation of $J_c$ quite difficult, due to the uncertainties in determining the current carrying length within the sample.

The first observation of a peak effect in this family of compounds is also reported. In analogy to the behaviour observed in high $T_c$ cuprates, this could indicate a 2D-3D crossover of the vortex lattice.

**Acknowledgements**

We acknowledge stimulating discussions with R. Lortz, R. Viennois and M. Therasse. We also thank R. Modoux and A. Ferreira for their technical support and M. Moret for the help in the design of the specific heat probe. This work was supported by the Swiss National Science Foundation through the National Centre of Competence in Research “Materials with Novel Electronic Properties” (MaNEP).

## Figure captions

**Figure 1** (a) X-ray diffraction pattern for the sample with nominal composition $SmFeAsO_{0.85}F_{0.15}$; SmOF impurity phase is indicated with an asterisk. (b) Scanning electron microscope images for the $SmFeAsO_{0.85}F_{0.15}$ polycrystalline sample, showing conglomerate particle sizes varying between 5 and 30 μm.

**Figure 2** $C/T$ versus $T$ at zero field (black squares) and 20 T (grey circles). Inset (a): superconducting transition from AC susceptibility. Inset (b): low temperature specific heat anomaly, related to the magnetic ordering of the $Sm^{3+}$ ions [17].

**Figure 3** Superconducting contribution to the specific heat for $B$ = 0, 6, 10, 14 and 20 T. The temperatures corresponding to $T_{mid}$, the midpoint of the calorimetric transition and to $T_{max}$, the maximum of $C/T$ at the superconducting jump are indicated for the curve at zero field.

**Figure 4** Distribution of $T_c$ at zero field and at $B$ = 10 T, obtained by the deconvolution of the calorimetric data.

**Figure 5** Upper critical field $B_{c2}$ versus $T/T_c$ for the $SmFeAsO_{0.85}F_{0.15}$ sample as determined from the midpoint $T_{mid}$ (solid diamonds) and the maximum $T_{max}$ (open diamonds) of the superconducting transition in the specific heat curves. The $B_{c2}$ values for $LaFeAsO_{1-x}F_x$ in Ref. [24] and [25] are also reported for comparison.

**Figure 6** Magnetic moment as a function of the applied magnetic field $B$ at $T$ = 5 K (solid squares) and 55 K (open circles). The $M(B)$ curve at 5 K is the superposition of a superconducting contribution and a ferromagnetic background.

**Figure 7** Superconducting contribution to the magnetic moment as a function of $B$, at $T$ = 5 K (a) and 25 K (b). The superconducting signal was isolated from the ferromagnetic background determined above $T_c$.

**Figure 8** Real part of the AC susceptibility $\chi'$ versus $T$: the double transition in the curve is determined by the crossover from intragrain to intergrain superconductivity, as a consequence of the granularity of the sample.

**Figure 9** (a) FE-SEM image showing grain sizes varying between less than 0.1 μm and 2-3 μm. (b) Magnification of a submicrometric grain.

**Figure 10** Inductive critical current density $J_c$ versus $B$ from 5 to 40 K. $J_c$ has been determined for $\langle R \rangle$ = 10 μm (left axis) and $\langle R \rangle$ = 0.1 μm (right axis), the current carrying length scale within the sample being not measured yet.

**Figure 11** *B*-*T* phase diagram for $SmFeAsO_{0.85}F_{0.15}$. Data points in the $B_{c2}(T)$ curve were determined from specific heat measurements. The irreversibility line and the $B_{peak}(T)$ line were extracted from magnetization measurements. The solid line is a fit $B_{irr} \propto [1-(T/T_c)]^2$.

Figure 1a

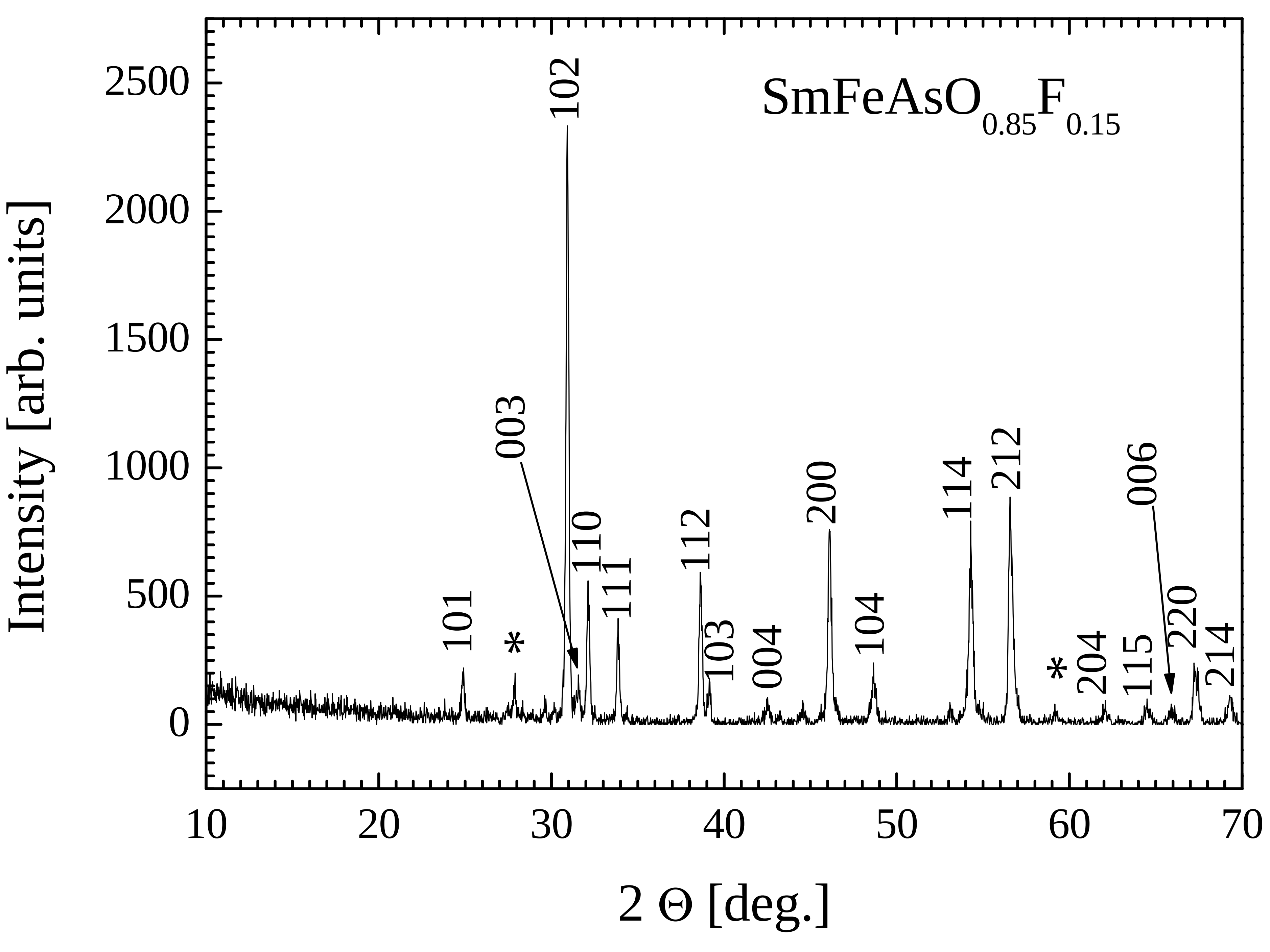

Figure 1b

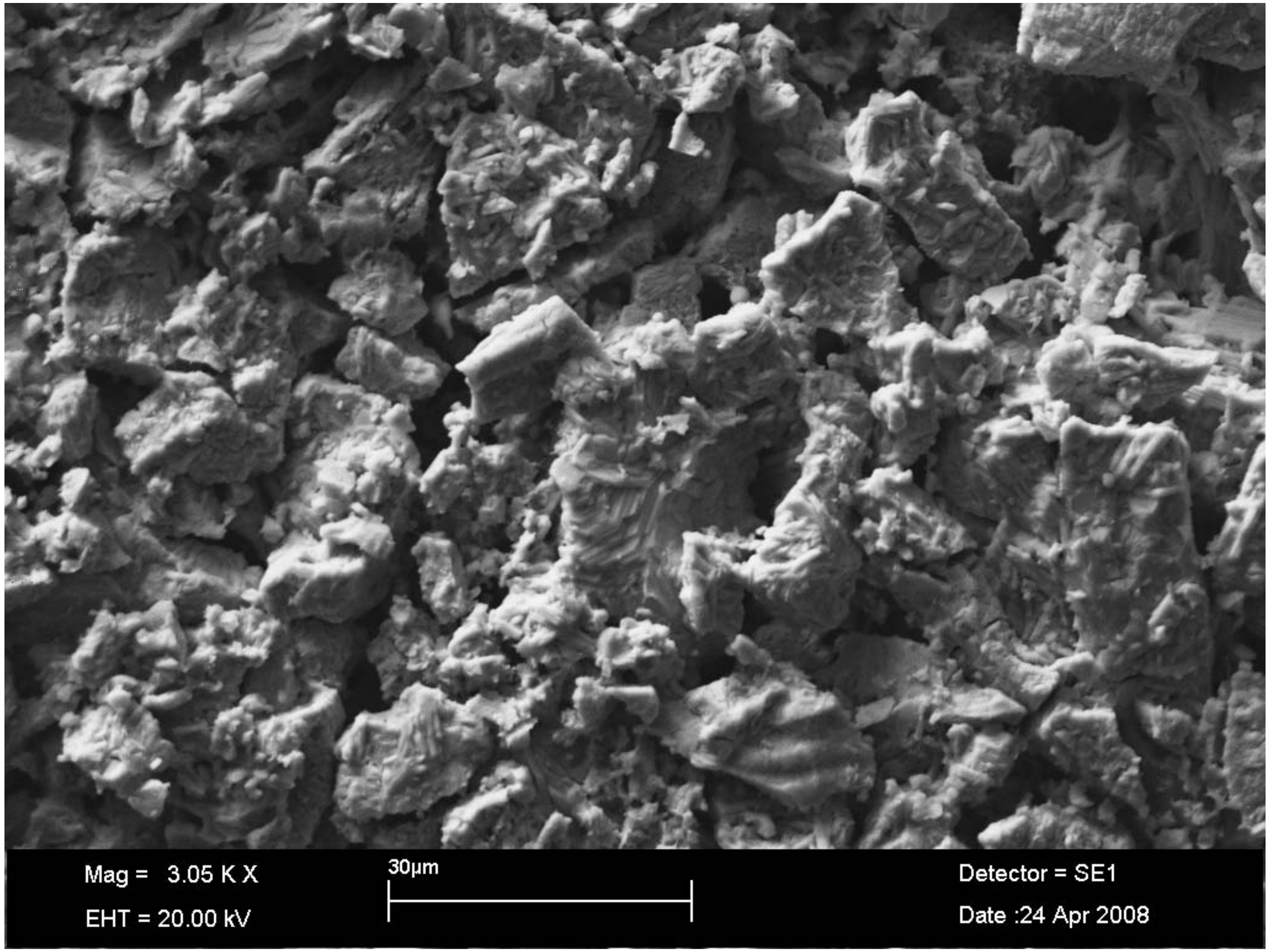

Figure 2

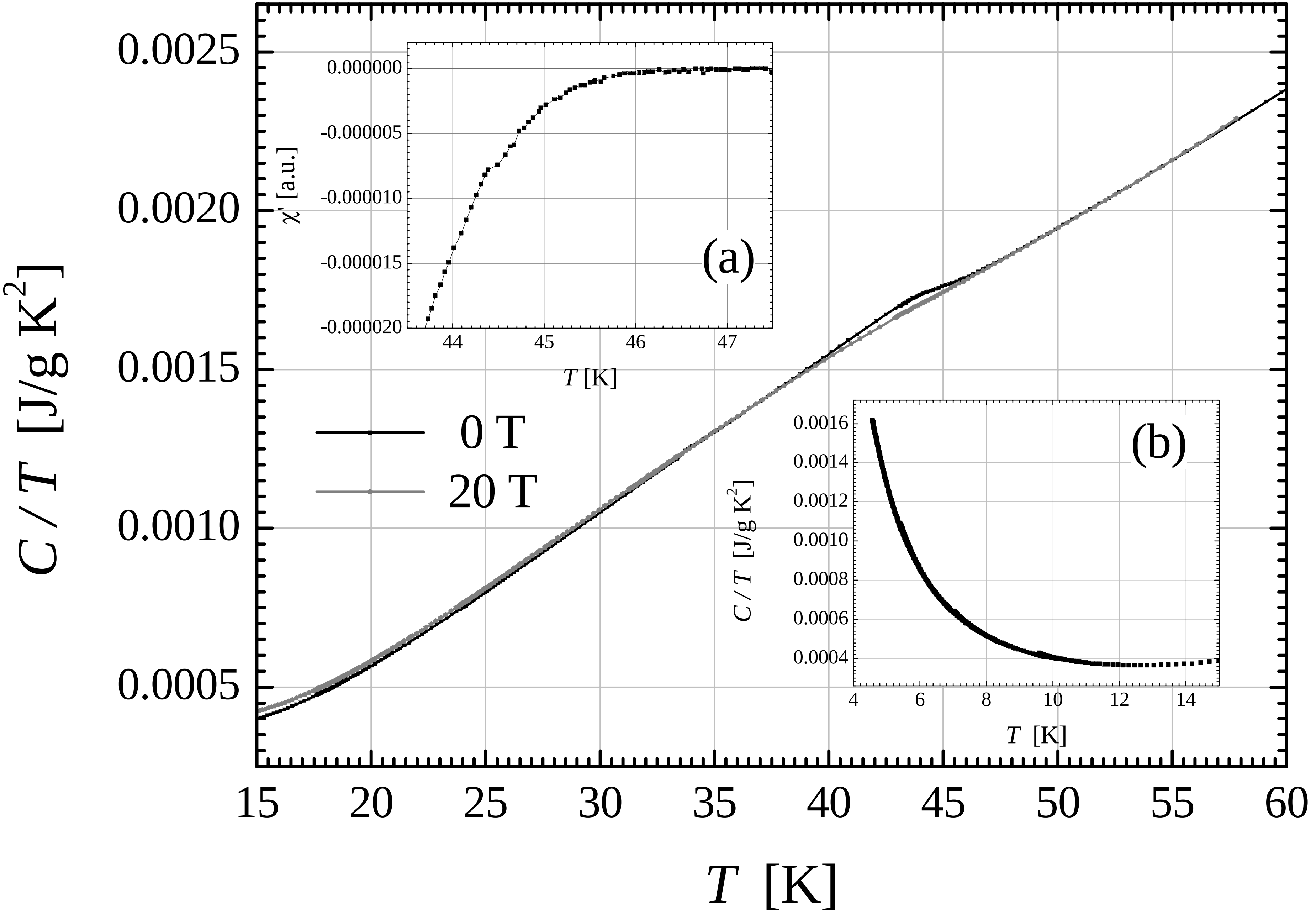

Figure 3

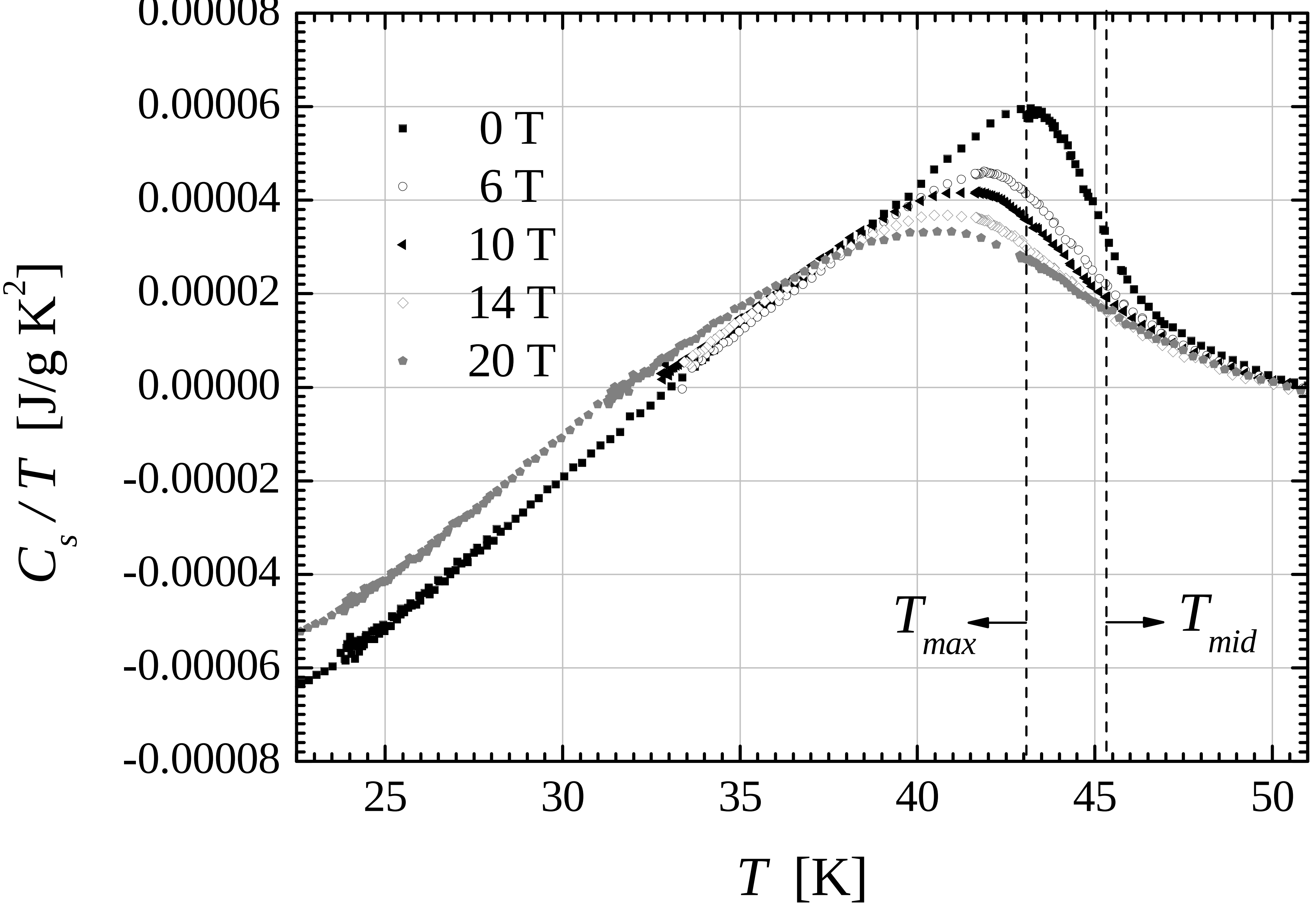

Figure 4

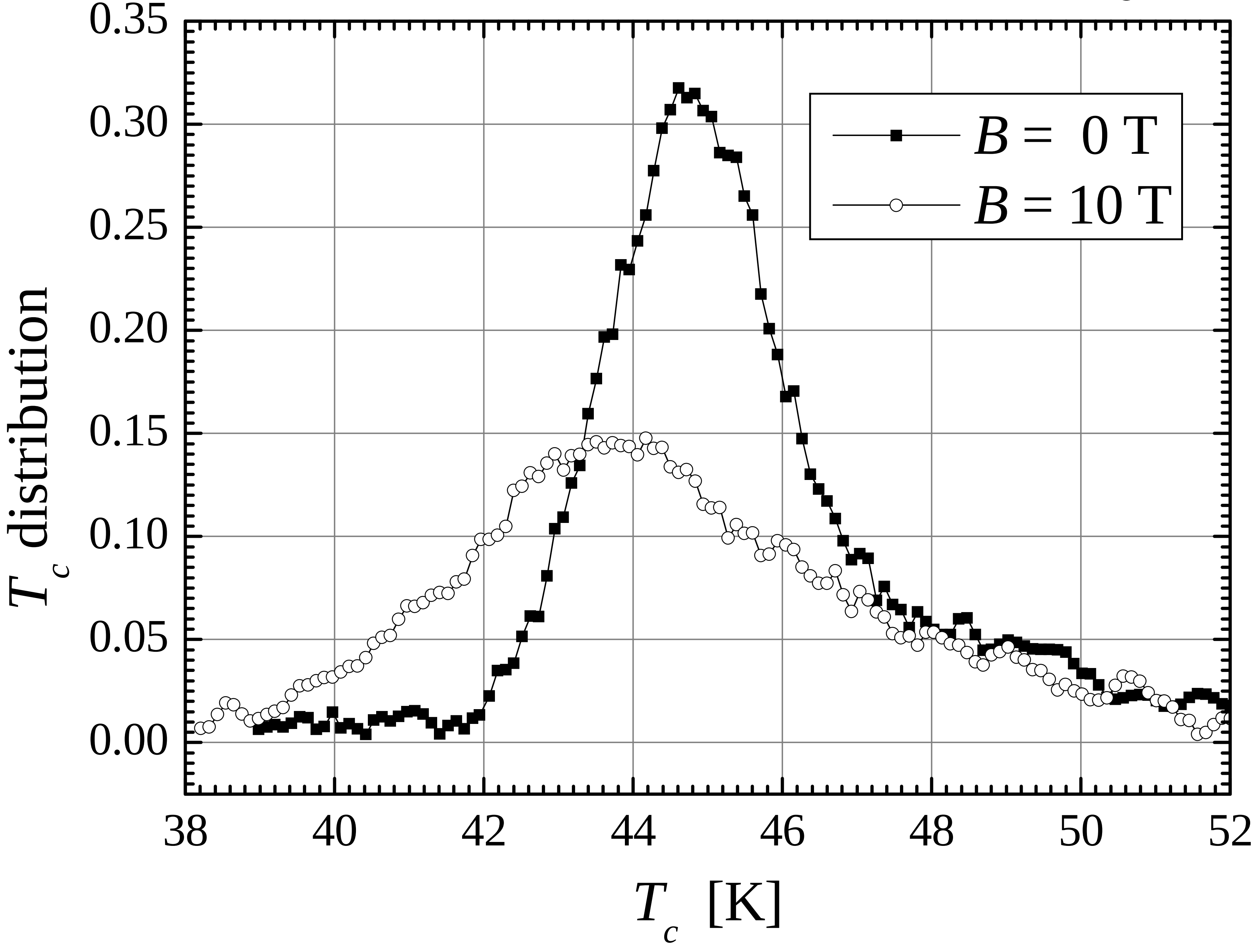

Figure 5

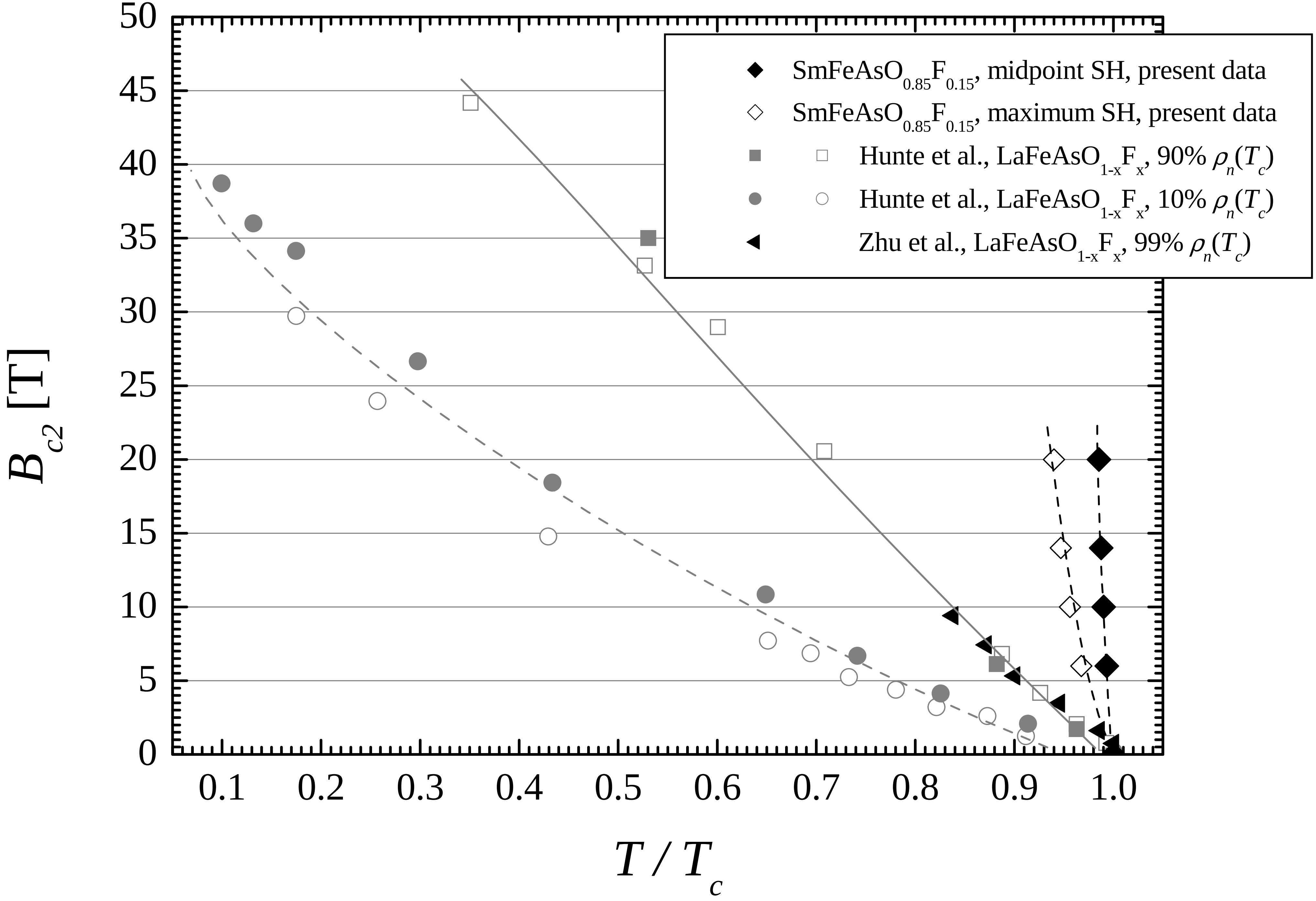

Figure 6

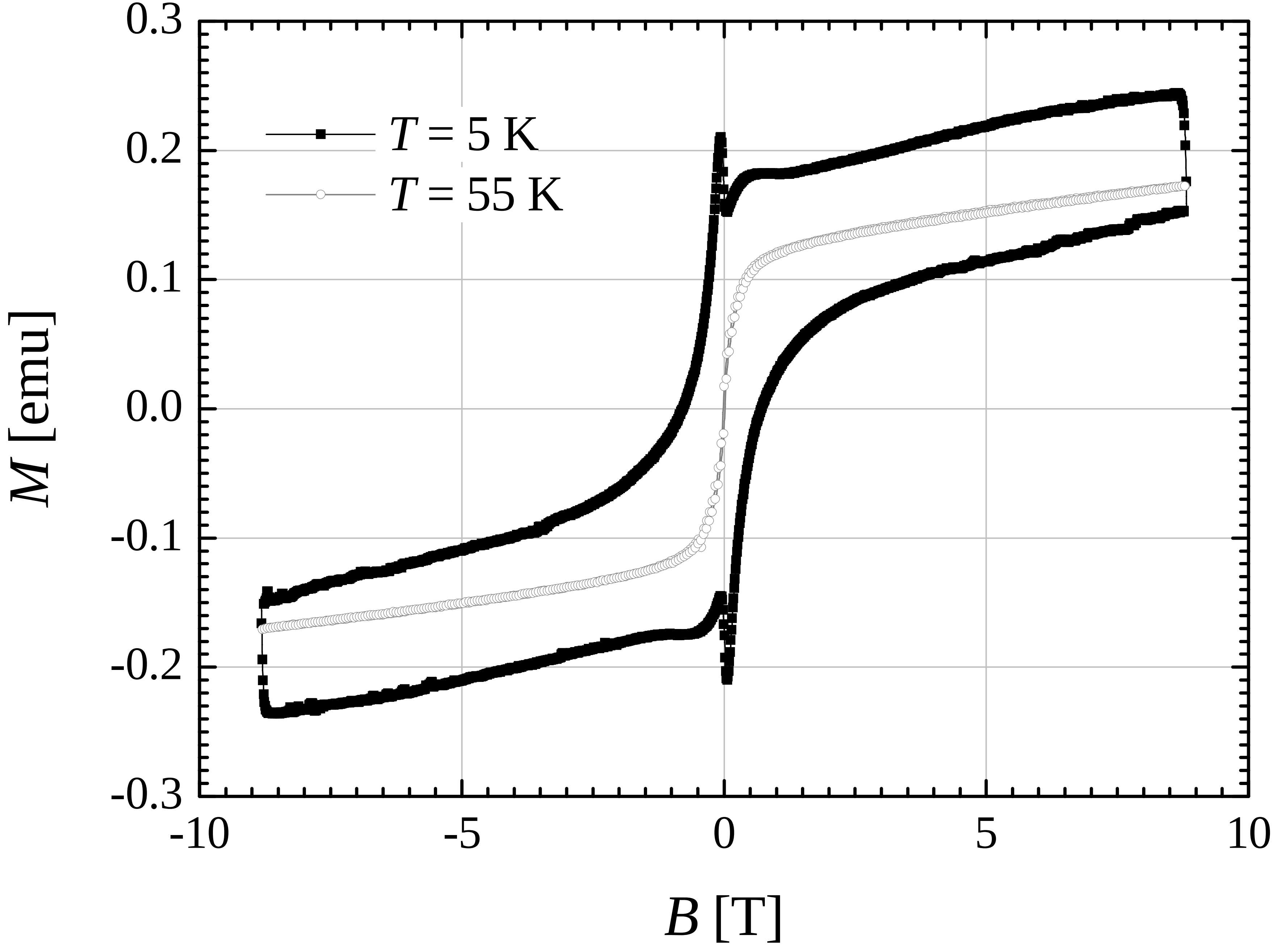

Figure 7a

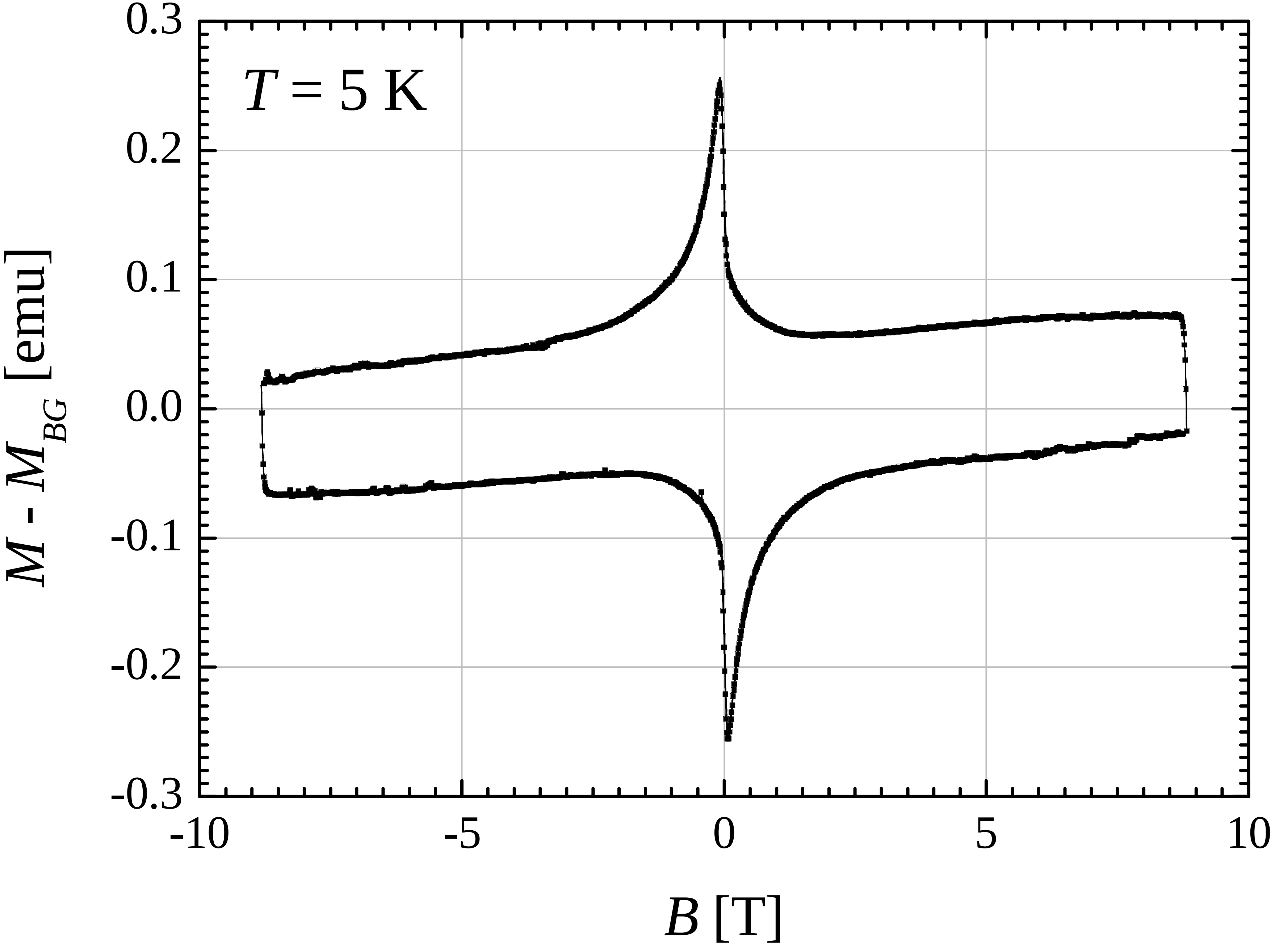

Figure 7b

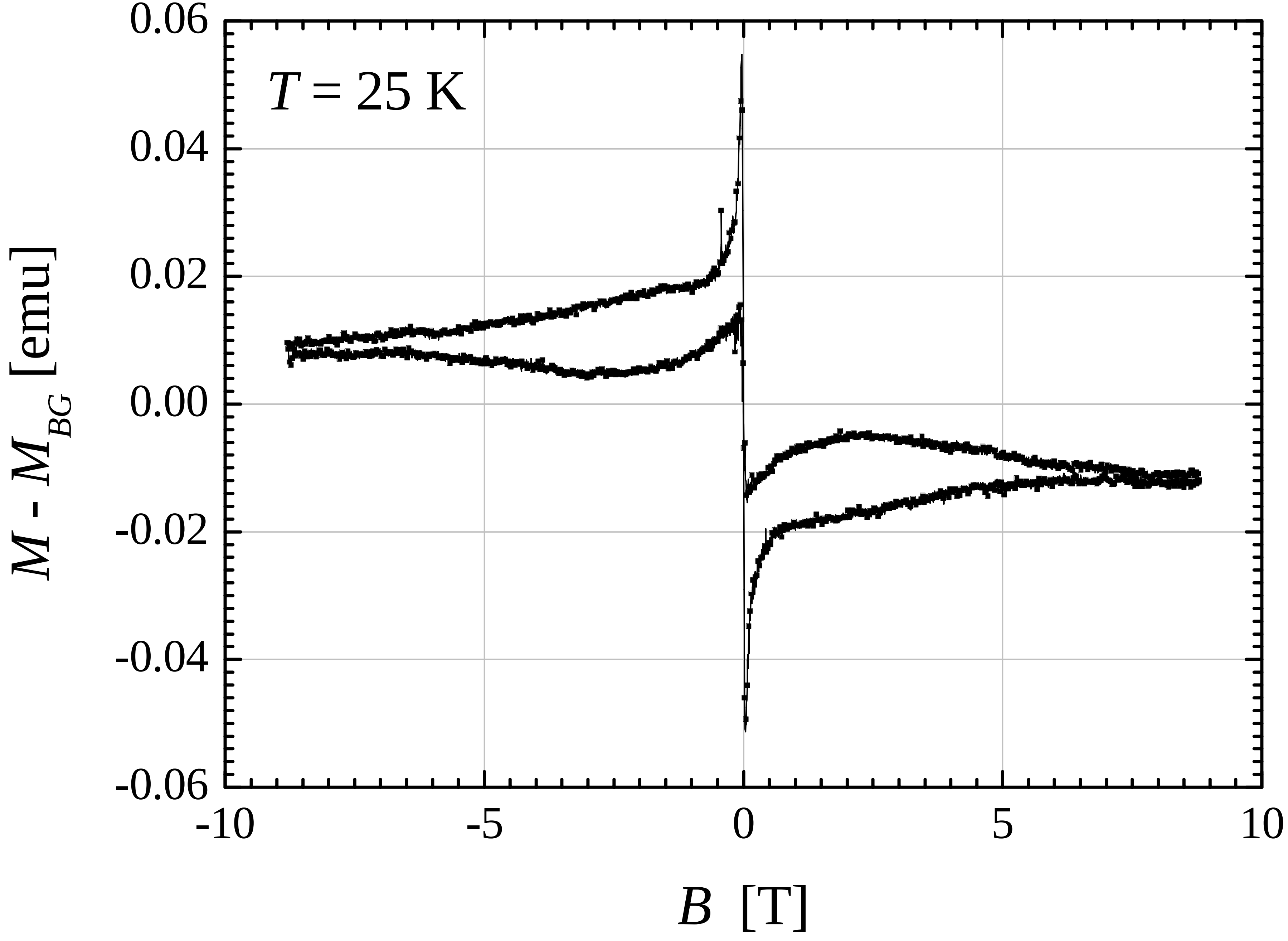

Figure 8

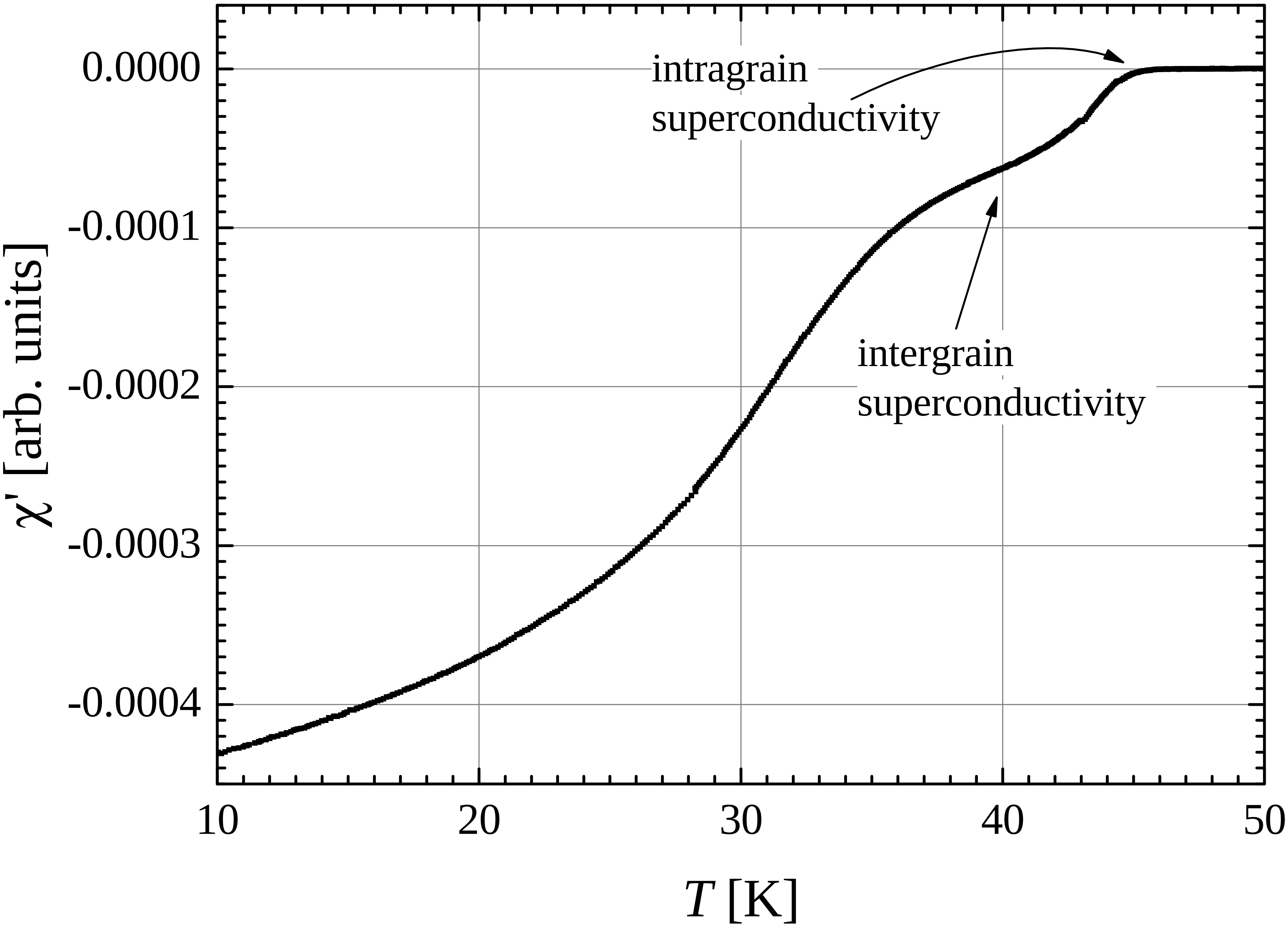

Figure 9a

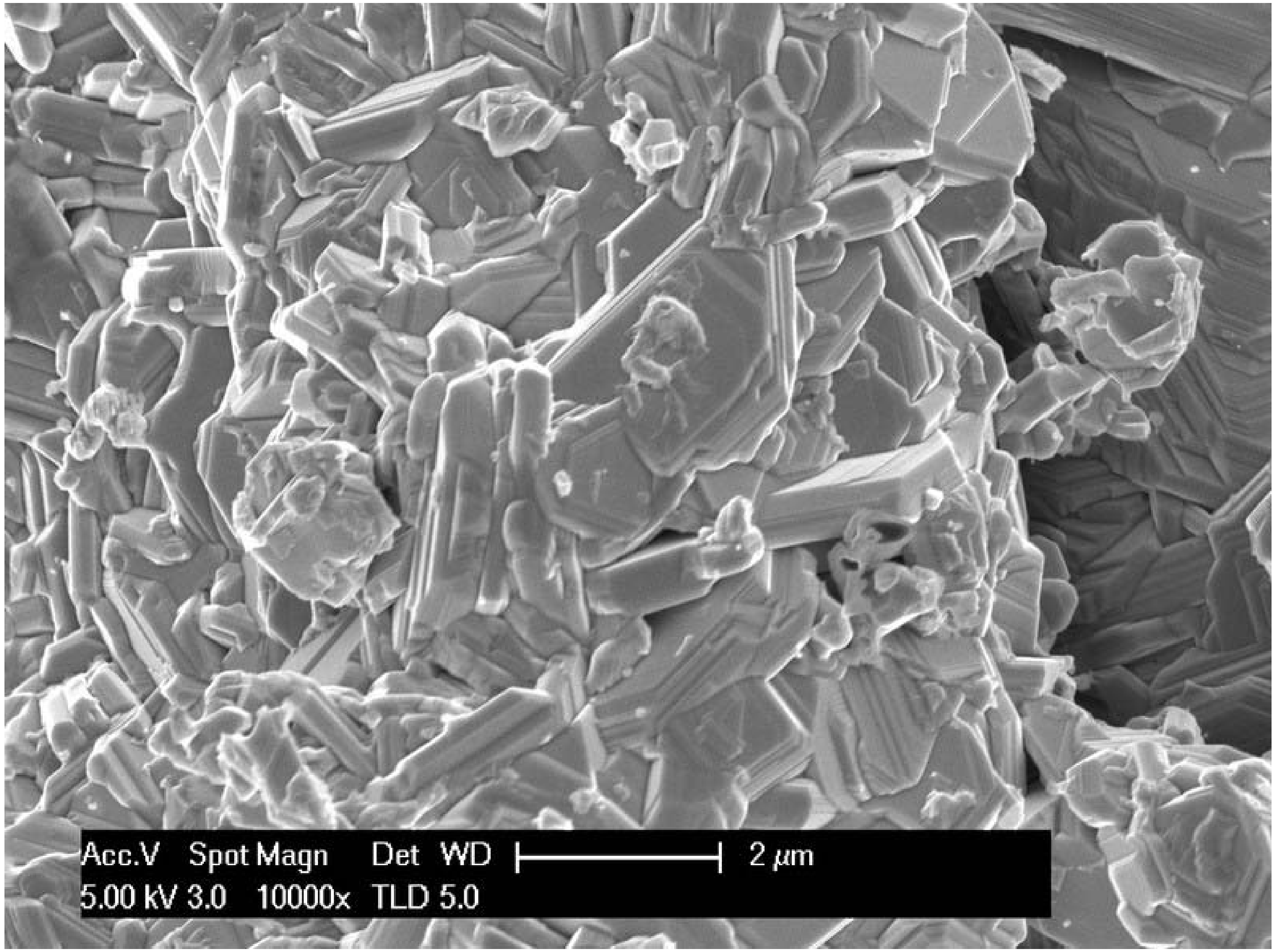

Figure 9b

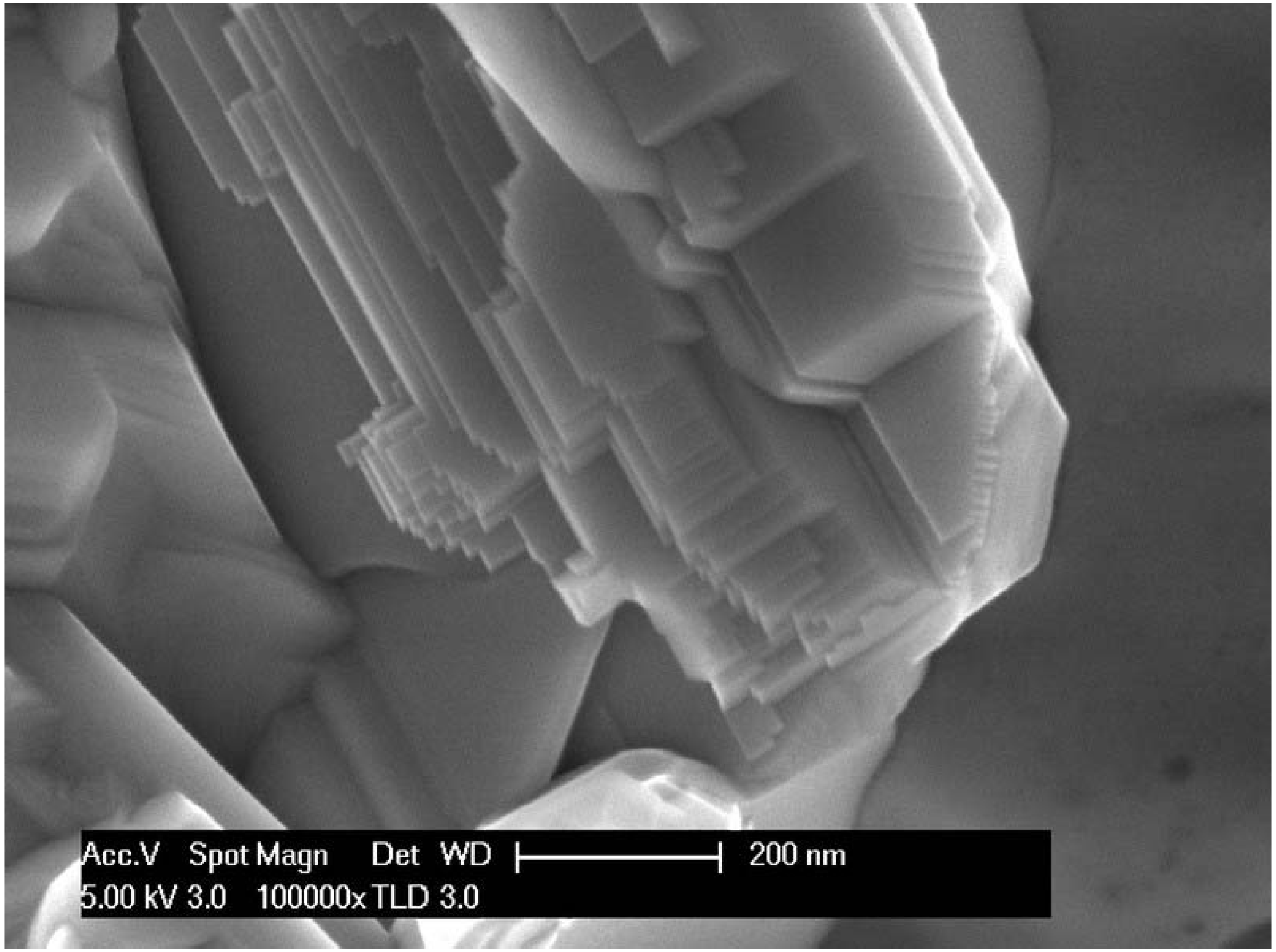

Figure 10

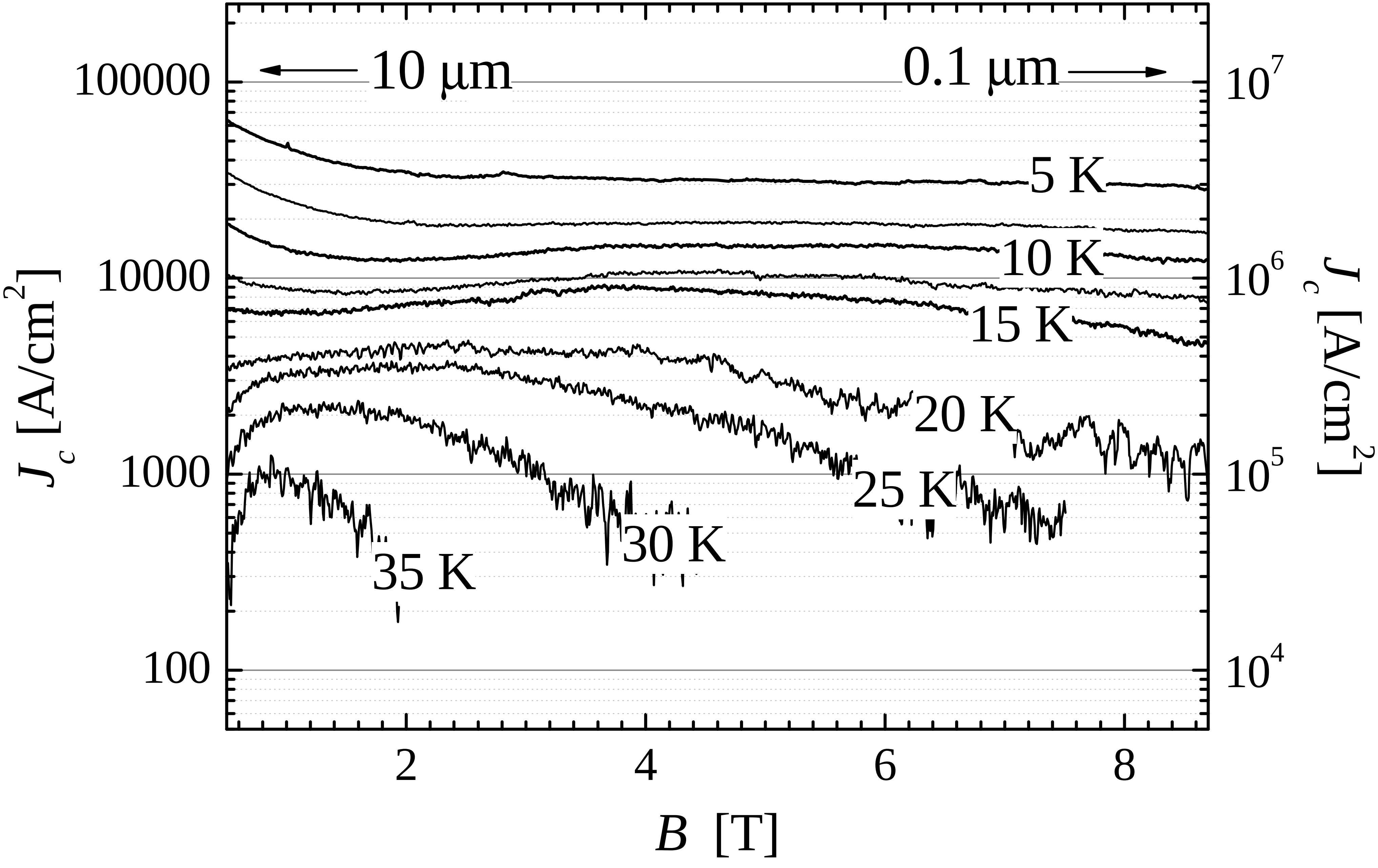

Figure 11

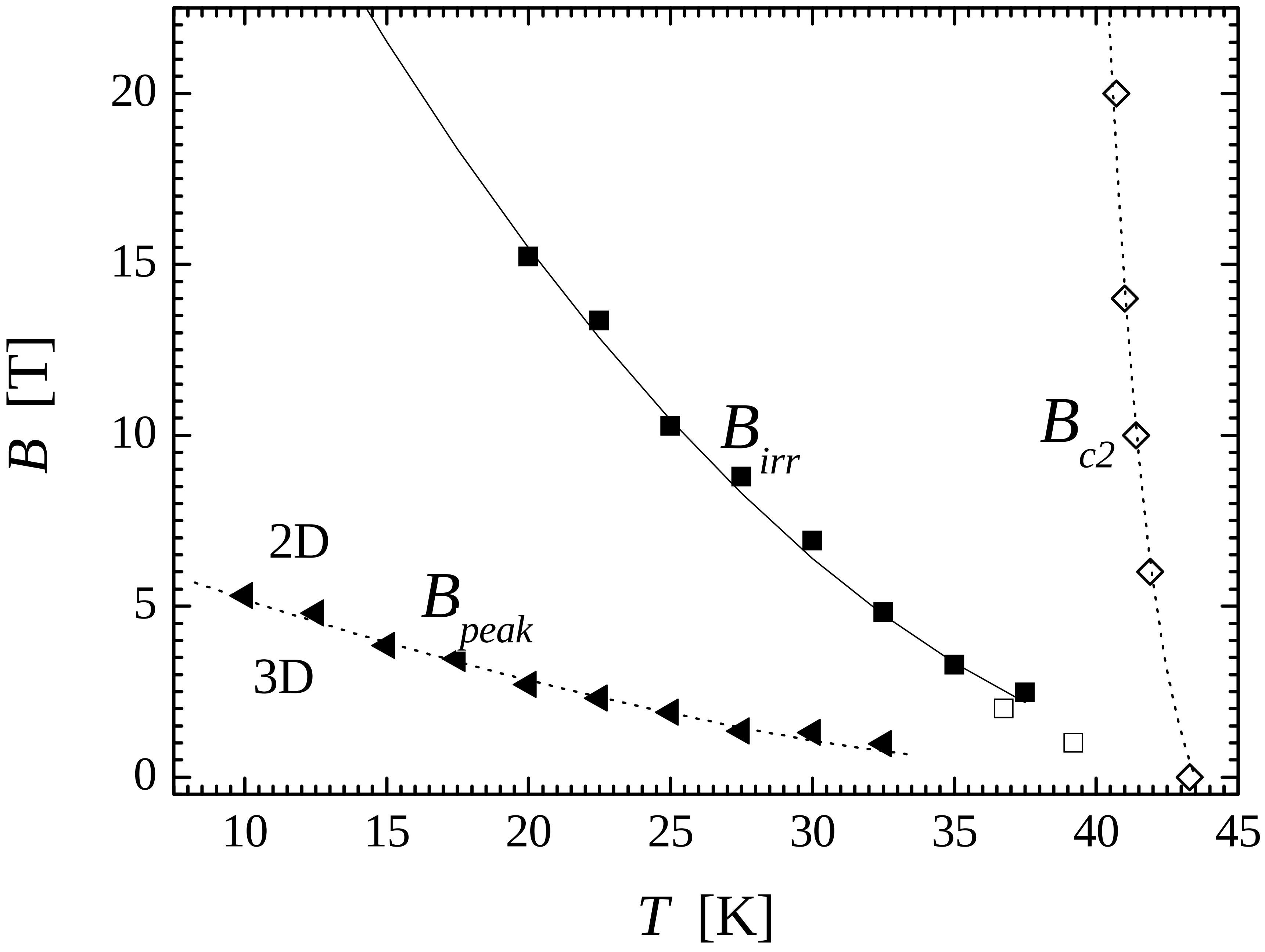